\documentclass{raa}           

\usepackage{graphicx,times}
\usepackage{natbib}
\usepackage{amssymb,amsmath}
\usepackage{CJK}
\bibpunct{(}{)}{;}{a}{}{,}

\usepackage[pagebackref=true]{hyperref}
\hypersetup{colorlinks = true, linkcolor = green, anchorcolor = red, citecolor = blue, filecolor = red, pagecolor = red, urlcolor = red}

\begin{document}
\begin{CJK*}{UTF8}{gbsn}
   \title{New Massive Contact Twin Binary in a Radio-quiet \ion{H}{ii} Region Associated with the M17 Complex}

 \volnopage{ {\bf 2022} Vol.\ {\bf 22} No. {\bf 3}, 1674-4527}
   \setcounter{page}{1}
   \author{Jia Yin (尹佳)
   \inst{1,2}, Zhiwei Chen (陈志维)\inst{3}, Yongqiang Yao (姚永强) \inst{1}, Jian Chen (陈健)\inst{3,4}, Bin Li (李彬)\inst{3,4}, and Zhibo Jiang (江治波)\inst{3} 
   }

   \institute{ National Astronomical Observatories,
Chinese Academy of Sciences,
Beijing 100101, China\\
        \and
             University of Chinese Academy of Sciences,
Beijing 100049, China\\
	\and
Purple Mountain Observatory,
Chinese Academy of Sciences,
Nanjing 210023, China; {\it zwchen@pmo.ac.cn}\\ 
\and
             School of Astronomy and Space Science, University of Science and Technology of China, Hefei 230026, China\\
\vs \no
   {\small Received 2021 December 12; accepted 2022 January 5; published 2022 February 25}
}
\abstract{Early-B stars, much less energetic than O stars, may create an \ion{H}{ii} region that appears as radio-quiet. We report the identification of new early-B stars associated with the radio-quiet \ion{H}{ii} region G014.645--00.606 in the M17 complex. The ratio-quiet \ion{H}{ii} region G014.645--00.606 is adjacent to three radio-quiet WISE \ion{H}{ii} region candidates. The ionizing sources of the radio-quiet \ion{H}{ii} regions are expected to later than B1V, given the sensitivity about $1-2$ mJy of the MAGPIS 20 cm survey. The stars were first selected if their parallaxes of Gaia EDR3 match that of the 22 GHz H$_2$O maser source within the same region. We used the color-magnitude diagram made from the Zwicky Transient Facility photometric catalog to select the candidates for massive stars because the intrinsic $g-r$ colors of massive stars change little from B-type to O-type stars. Five stars lie in the areas of the color-magnitude diagram where either reddened massive stars or evolved post-main sequence stars of lower masses are commonly found. Three of the five stars, sources 1, 2, and 3, are located at the cavities of the three IR bubbles, and extended H$\alpha$ emission is detected around the three IR bubbles. We suggest that sources 1, 2, and 3 are candidates for early-B stars associated with the radio-quiet region G014.645--00.606. Particularly, source 1 is an EW type eclipsing binary with a short period of 0.825 day, while source 2 is an EA type eclipsing binary with a short period of 0.919 day. The physical parameters of the two binary systems have been derived through the PHOEBE model. Source 1 is a twin binary of two stars with $T_\mathrm{eff}\approx23,500\,\mathrm{K}$, and source 2 contains a hotter component ($T_\mathrm{eff}\approx20,100\,\mathrm{K}$) and a cooler one ($T_\mathrm{eff}\approx15,500\,\mathrm{K}$). The $O-C$ values of source 1 show a trend of decline, implying that the period of source is deceasing. Source 1 is likely a contact early-B twin binary, for which mass transfer might cause its orbit to shrink. 
}

   \authorrunning{J. Yin et al. }            
   \titlerunning{The Massive Contacting Twin Binary in M17}  
   \maketitle

\section{Introduction}           
\label{sec:intro}

Our understanding of the OB stars (mass $>8\,M_{\sun}$) is far from enough and precise. For instance, systematic surveys of the OB stars in the past decade have shown that the majority of OB stars are binaries and/or multiple systems \citep{2012Sci...337..444S,2012MNRAS.424.1925C,2014ApJS..211...10S,2014ApJS..213...34K,2015AJ....149...26A,2019MNRAS.490.5147P,2021A&A...652A.120I,2021RAA....21..272L}. The binary fraction of OB stars increases as the spectral type goes to early \citep{2022RAA....22b5009G}, and O stars in cluster, associations, and runaways have binary fraction (70\%) higher than those in field environment (40\%; \citealt{2012MNRAS.424.1925C}). These observational studies challenge the accepted predominance of the single star evolutionary channel and complicate the exact determination of the evolutionary status and final fates of these stars \citep{2020RAA....20..161H}.

OB stars are mostly within stellar clusters, thus the dynamical interactions between cluster members likely alter the properties of OB-type multiple systems after several dynamical times \citep{2015A&A...582A..42K}. Establishing the multiplicity properties of large, statistically significant samples of newly born OB stars is consequently crucial to understand the initial conditions and properly compute the evolution of these objects. \citet{2022A&A...658A..69B} analyzed the spectrum of 80 B-type stars in the open cluster NGC 6231 with an age of 2--7 Myr, and constrained a de-biased spectroscopic binary fraction of $52\%\pm8\%$, a value lower than that for the more massive O-type stars. The orbital properties of the 27 B-type binary candidates in NGC 6231 generally resemble those of B- and O-type stars in both the Galaxy and Large Magellanic \citep{2022A&A...658A..69B}. Spectroscopic time domain observations for newly born OB stars distributed in various star-forming regions are generally time consuming. For the star-forming regions, multiband photometric surveys of the young stars from low to high-mass are the efficient ways to detect young Eclipsing Binaries (EBs; \citealt{2013A&A...557A..13B,2015AJ....150..132R,2019MNRAS.487.3505M,2019A&A...627A.135B,2021AJ....162...52Y,2021AJ....161..120M}). The physical parameters of multiple systems can be determined by modeling the light and radial velocity curves, and provide a detailed understanding of their interior rotation and angular momentum transport mechanisms \citep{2019ARA&A..57...35A,2021MNRAS.501L..65S}. 

Apart from the optical spectroscopic surveys for the OB stars in the Galaxy \citep{2011ApJS..193...24S,2015AJ....149...26A,2019ApJS..241...32L,2021ApJS..253...54L}, the mid-IR surveys are able to detect the mid-IR radiation from warm dust grains and polycyclic aromatic hydrocarbons (PAH) which interplay with the (far)-UV photons from OB stars \citep{2014ApJS..212....1A,2015ApJ...799..153K,2019MNRAS.488.1141J}. The ionized gas excited by far-UV photons from OB stars is also detectable in the form of free-free continuum emission and radio recombination lines in radio wavelengths \citep{2006AJ....131.2525H,2006AJ....132.1158S,2011ApJS..194...32A,2018ApJS..234...33A,2019MNRAS.489.4862H,2021ApJS..254...36W,2021RAA....21..209Z}. The methods from mid-IR and radio wavelengths cannot constrain the specific candidate OB stars responsible for the mid-IR and radiation emissions. In addition, photometric and astrometric information of the stellar sources within the same areas are necessary for further classifying the candidate OB stars. For example, new young OB stars were found in the field around young clusters from the VST Photometric H${\alpha}$ Survey of the Southern Galactic Plane and Bulge (VPHAS+; \citealt{2014MNRAS.440.2036D,2017MNRAS.465.1807M,2018MNRAS.480.2109D}). Star-forming regions are generally suffering high extinction, OB stars within star-forming regions are largely unknown and awaiting discovery \citep{2015A&A...578A..82C,2017ApJ...838...61P}.

The M17 \ion{H}{ii} nebula (aka NGC 6618) is located in the Sagittarius spiral arm \citep{1979ApJ...230..415E,2019ApJ...885..131R} at a distance $\approx$2.0\,kpc \citep{2011ApJ...733...25X,2016MNRAS.460.1839C}. The young embedded cluster NGC 6618 is thought to be younger than 3 Myr \citep{2002ApJ...577..245J}, and contains hundreds of stars earlier than B9 \citep{1991ApJ...368..432L,2008ApJ...686..310H}. M17 has been studied in depth from various aspects, such as physical properties of massive young stellar objects (MYSOs; \citealt{2004Natur.429..155C,2007ApJ...656L..81N,2017A&A...604A..78R,2020ApJ...888...98L,2021ApJ...922...90C}), magnetic fields of the \ion{H}{ii} region and surrounding molecular clouds \citep{2012PASJ...64..110C}, the molecular gas distribution and overall star formation activities \citep{2009ApJ...696.1278P,2020ApJ...891...66N}. As the mid-IR surveys like GLIMPSE and MIPSGAL covered the Galactic plane, an infrared dark cloud (IRDC) located southwest to M17 came into common interest. With the name M17 SWex, this IRDC was first studied by \citet{2010ApJ...714L.285P} who identified numerous candidate YSOs within M17 SWex \citep[see also][]{2016ApJ...825..125P}. The large-scale CO gas of M17 and M17 SWex share common molecular gas radial velocity \citep{1979ApJ...230..415E,2020ApJ...891...66N}, indicating that M17 and M17 SWex are parts of the entire M17 cloud complex. \citet{2019PASJ...71S...6S} observed M17 SWex with the 45\,m Nobeyama radio telescope in $^{12}$CO, $^{13}$CO, C$^{18}$O, CCS, and N$_2$H+ emission lines in $93-115$ GHz, and detected 46 dense cores (mostly gravitationally stable) within M17 SWex. Two extended green objects (EGO) $G14.33-00.64$ and $G14.64-00.57$ were detected toward M17 SWex \citep{2019ApJ...875..135T}. The EGO $G14.33-00.64$ is at a distance $1.12\pm0.13$ kpc \citep{2010PASJ...62..287S}, while EGO $G14.64-00.57$ is at a distance $1.83^{+0.08}_{-0.07}$ kpc \citep{2014A&A...566A..17W}. The molecular gas associated with EGO $G14.64-00.57$ and the bulk gas of M17 are coherent in $V_{LSR} \approx 19\,\mathrm{km\,s^{-1}}$ \citep{2019PASJ...71S...6S,2020ApJ...891...66N}. EGO $G14.64-00.57$ is associated with both dense core 46 in \citet{2019PASJ...71S...6S} and hot core candidate $G14.630-0.569$ in \citet{2021PASJ...73..568S}. These signposts for EGO $G14.64-00.57$ may hint earlier stage of high-mass star formation \citep{2019ApJ...875..135T} or formation of less massive stars resulting in radio-quiet infrared sources \citep{2014ApJS..212....1A}. Not far from EGO $G14.64-00.57$, three bright WISE sources are classified as radio-quiet \ion{H}{ii} region candidates \citep{2014ApJS..212....1A}. These sources are also coinciding with an IR bubble \citet{2019MNRAS.488.1141J}. The IR brightness and morphology hint the existence of embedded massive stars.  

In this paper we report the discoveries of new OB star candidates that are still embedded within the molecular gas of the M17 cloud complex. Two candidates are identified as short-period EBs from the multiepoch and multiband photometric observations. This paper is organized as follows: Section~\ref{sec:data} describes the observations and uses public data; in Section~\ref{sec:h2}, the new radio-quiet \ion{H}{ii} region in M17 SWex is presented; in Section~\ref{sec:candidates}, the most probable OB candidates associated with the \ion{H}{ii} are identified; in Section~\ref{sec:obvar}, the variabilities and parameters of OB binaries are analyzed. Finally, our conclusions are summarized in Section~\ref{sec:con}.

\section{Observations and Public Data}
\label{sec:data}
\subsection{The UKIRT near-IR Imaging Data}
The UKIRT/WFCAM $JHK$ observations of the M17 \ion{H}{ii} region were taken by the UKIRT telescope with the WFCAM near-IR camera on 2005 June 01-04 under the project U/05A/J2, and the H$_2$ $1-0$ S(1) narrow filter observations were taken by the same camera on 2008 May 21 and 23 under the project U/08A/H59. The reduced $JHK$ and H$_2$ images were directly retrieved from the WFCAM Science Archive\footnote{\url{http://wsa.roe.ac.uk/index.html}}. The total exposure time of the $JHK$ and H$_2$ images is 2520\,s, 2160\,s, 2160\,s, and 800\,s, respectively.

\subsection{High-cadence Optical Observations}
We carried out high-cadence time-series photometric observations of the \ion{H}{ii} region candidates in M17, using the 0.8\,m Yaoan high precision telescope (hereafter YAHPT) at Purple Mountain Observatory. The YAHPT is equipped with a $2048\times2048$ CCD and the field of view is about $11\arcmin\times11\arcmin$. The Johnson-Cousions filters $V$, $R_c$, and $I_c$ were used and the exposure times are all 30\,s in these filters. Multiband images were obtained on 11 nights with good weather from 2021 July 7 to August 7. Aperture photometry was performed on the multiband images obtained in these nights. 
 
\subsection{Public Survey Data}
The Gaia Early Data Release 3 (Gaia EDR3; \citealt{2021A&A...649A...1G}) provides astrometric solutions for about 1.3 billion stars in the Galaxy. The limiting magnitude of Gaia EDR3 is about 21\,mag. The typical uncertainties of the positions, parallaxes, and proper motions at $G=20$\,mag are 0.4\,mas, 0.5\,mas, and 0.5\,mas yr$^{-1}$, respectively. 

The Zwicky Transient Facility (ZTF; \citealt{2019PASP..131a8002B,2019PASP..131a8003M}) is a northern-sky synoptic survey using the Palomar 48-inch Schmidt Telescope. Aided by the wide-field camera with a FOV of 47 deg$^2$, ZTF is able to scan the entire northern sky in two nights, and to achieve a depth of 20.5 mag with 30 s exposure time in the three custom-made filters $g_{\rm ZTF}$, $r_{\rm ZTF}$, and $i_{\rm ZTF}$. The multiepoch photometric data of ZTF in the $g_{\rm ZTF}$ and $r_{\rm ZTF}$ filters were used in this paper.

The SuperCOSMOS H$\alpha$ Survey (SHS; \citealt{2005MNRAS.362..689P}) provides the online digital atlas of the AAO/UKST H$\alpha$ survey for the southern Galactic Plane and Magellanic Clouds. The narrow band H$\alpha$ image is a good tracer of diffuse ionized gas. 

The Galactic Legacy Infrared Mid-Plane Survey Extraordinaire (GLIMPSE) survey \citep{2003PASP..115..953B,2009PASP..121..213C} is a legacy science program of the Spitzer Space Telescope, and provides the inner Galactic plane survey in 3.6, 4.5, 5.8 and 8.0\,$\mu$m bands with angular resolution about $2\farcs5$. In the Spitzer $8\,\mu$m band, the PAH emission and warm dust emission are prominent. We used the Spitzer $8\,\mu$m image to trace the area in the vicinity of embedded OB stars.

\section{The new radio-quiet \ion{H}{ii} region in the southwest to M17}
\label{sec:h2}

\begin{figure}[!ht]
 \centering
          \includegraphics[width=.5\textwidth]{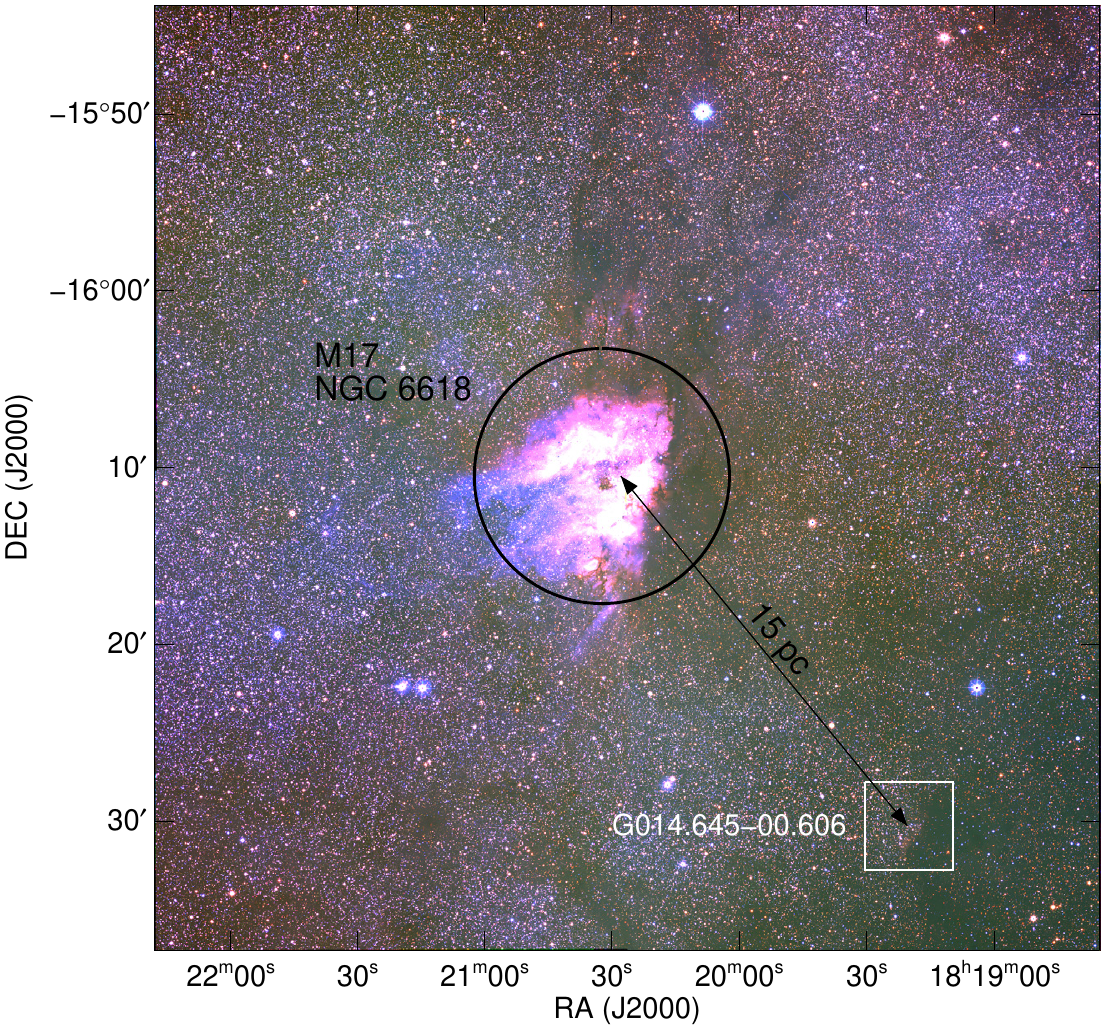}
          $\quad$
          \includegraphics[width=.45\textwidth]{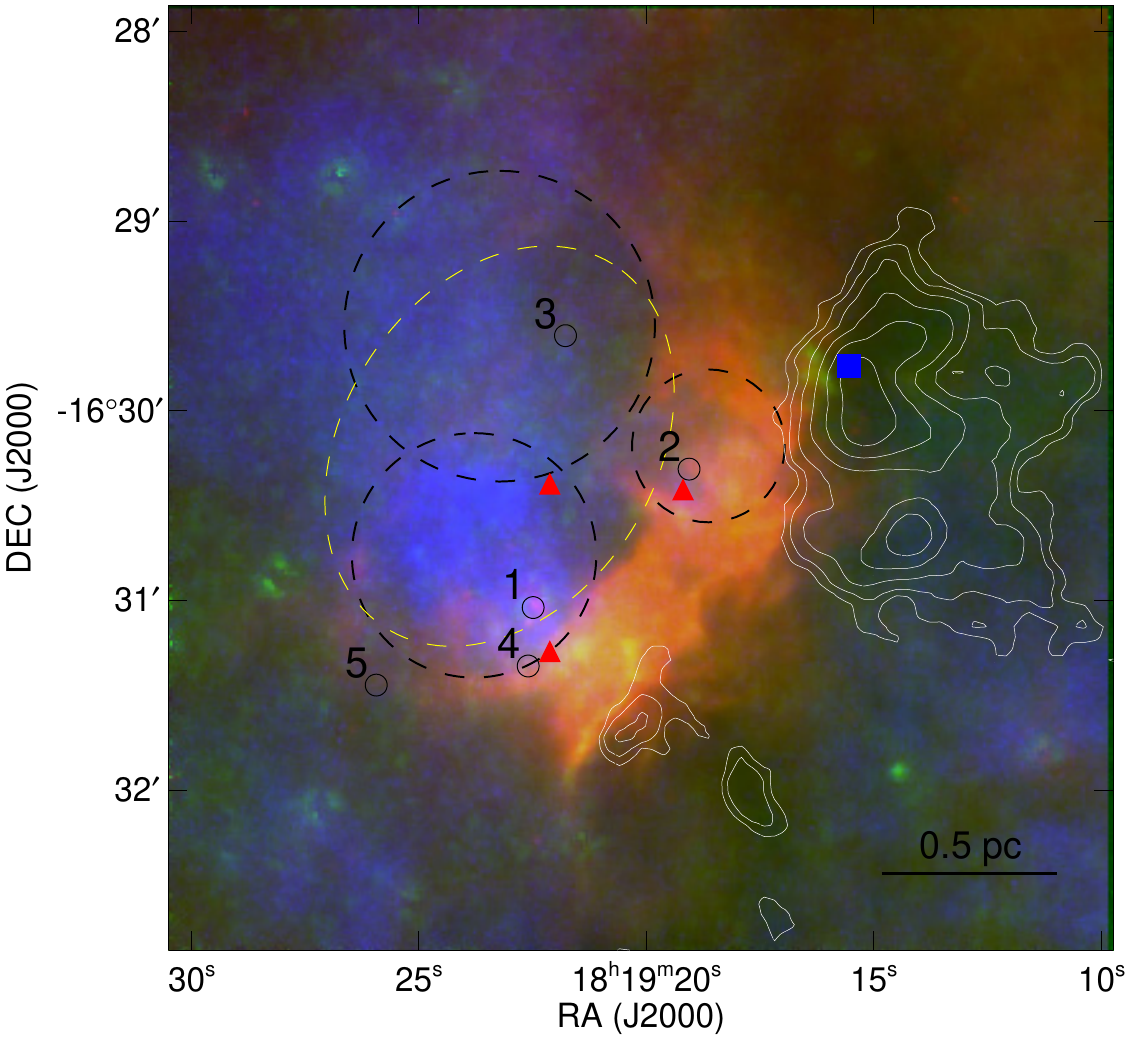}
      \caption{Color-composite image of M17 from UKIDSS (blue: $J$, green: $H$, and red: $K$) and G014.645--00.606 (blue: H$\alpha$, green: H$_2$, and red: 8\,$\mu$m). In the right panel, the cloud is overlaid with the 1.2 mm continuum emission contours \citep{2010A&A...515A..42R}. The dashed oval indicates IR bubble identified by \citet{2019MNRAS.488.1141J}. The three dashed circles indicate three circular bubbles in 8\,$\mu$m band. The blue square indicates the 22\,GHz water maser in the cloud \citep{1981ApJ...250..621J,1993A&AS..101..153P}. The black circles with IDs indicate the OB candidates. The red triangles indicate three WISE sources \citep{2014ApJS..212....1A}. The black line segment represents the view of 0.5\,pc at a distance of 1.83\,kpc.}\label{fig:m17}
\end{figure}

As shown in the left panel of Figure~\ref{fig:m17}, the $JHK$ color-composite image of 1 deg$^2$ area is centered on the bright M17 \ion{H}{ii} region. The majority of M17 SWex IRDC is located far to the southwest and thus is out of the 1 deg$^2$ area. In the lower right corner of the left panel of Figure~\ref{fig:m17}, an area of high extinction is obviously seen in the $JHK$ color-composite image. This high extinction area is corresponding to the most eastern part of the M17 SWex IRDC in the frame of galactic coordinate. The close-up view of this high extinction area is shown in the right panel of Figure~\ref{fig:m17}, which is composite of H$\alpha$ image, H$_2$ $2.12\,\mu$m image, and Spitzer $8.0\,\mu$m image. The EGO $G14.64-00.57$ is lying on the outer border of bright rim seen in the Spitzer $8.0\,\mu$m image, where the 1.2 mm continuum peak is coinciding with the molecular dense core 46 in \citet{2019PASJ...71S...6S} and hot core candidate $G14.630-0.569$ in \citet{2021PASJ...73..568S}.

The IR bubble identified in \citet{2019MNRAS.488.1141J} is shown as the dashed ellipse with a major axis of 1.5\,$\arcmin$. The IR bubble outlines the border of the bright rim seen in the Spitzer $8\,\mu$m image. A single star-driven bubble may not be able to form this complex profile, which may consist of multiple bubbles with different ages and distances. The Spitzer 8\,$\mu$m image clearly presents three smaller circular bubbles of radii less than 0.5\,pc, which are shown as black dashed-line circles. The circles are extended by the arched bright rims, and the bubble-1, bubble-2, and bubble-3 contains stellar source 1, source 2, and source 3 in line of sight, respectively. The IR elliptical bubble in \citet{2019MNRAS.488.1141J} seems to consist of the two larger circular bubbles (bubble-1 and bubble-3) outlined in this work. The smaller circular bubble (bubble-2) is within the $8.0\,\mu$m-bright rim. As shown in the red triangles, two of the three WISE \ion{H}{ii} region candidates in \citep{2014ApJS..212....1A} are lying on the border of the $8.0\,\mu$m-bright rim, with the third one within the IR elliptical bubble. Two WISE \ion{H}{ii} region candidates are coinciding with the edges of two circular bubbles (bubble-1 and bubble-2). Around the IR bubble, diffuse H$\alpha$ emissions extend to the northeast, in which direction little mid-IR emission is observed, indicative of little dust and gas. The H$\alpha$ emission within bubble-1 is very clear, while the H$\alpha$ emission around bubble-2 and bubble-3 is weak. The bubbles seen at 8\,$\mu$m and extended H$\alpha$ emission within the cavities of bubbles indicate the existence of massive stars. 

The three WISE \ion{H}{ii} candidates are classified as radio-quiet because of the absence of radio continuum sources \citep{2014ApJS..212....1A}. The sensitivity of the MAGPIS 20\,cm survey is 1-2 mJy \citep{2006AJ....131.2525H}. We assume an upper limit of 4 mJy at 20 cm for the three WISE \ion{H}{ii} candidates. From this upper limit of radio continuum emission, we can estimate the upper limit of the number of ionizing photons in this region. The number of ionizing photons was represented by \citet{1990ApJ...362..147C} as \begin{equation}
{N_\mathrm{LyC} = 9.0\times10^{43}(\frac{S_{\nu}}{\rm mJy})(\frac{d}{\rm kpc})^2(\frac{\nu}{5\,\rm GHz})^{0.1}\,{\rm photons\,s}^{-1}},
\end{equation}
where $S_{\nu}$ is the integrated flux density of radio continuum emission . $S_\nu\leqslant4$ mJy leads to N$_\mathrm{LyC}\leqslant1.0\times10^{45}\,\mathrm{photons\,s^{-1}}$ when $d=1.8\,$ kpc and $\nu = 1.4\,$GHz. According to the rates of Lyman and dissociating photons (Table\,1 in \citealt{1998ApJ...501..192D}), a ZAMS star of effective temperature about 23,000\,K could emit such amount of ionizing photons. The putative OB stars within the IR bubbles are not much hotter than 23,000\,K, or equivalently not earlier than B1V. 

This radio-quiet \ion{H}{ii} region candidate, named as G014.645--00.606, is located at the middle between the M17 \ion{H}{ii} region and M17SWex IRDC. In the following we will locate the massive star candidates of G014.645-00.606 based on the multiband photometric data and Gaia EDR3 astrometric data.

\section{Massive Star Candidates of G014.645--00.606}
\label{sec:candidates}

\begin{figure}[!ht]
 \centering
          \includegraphics[width=1.0\textwidth]{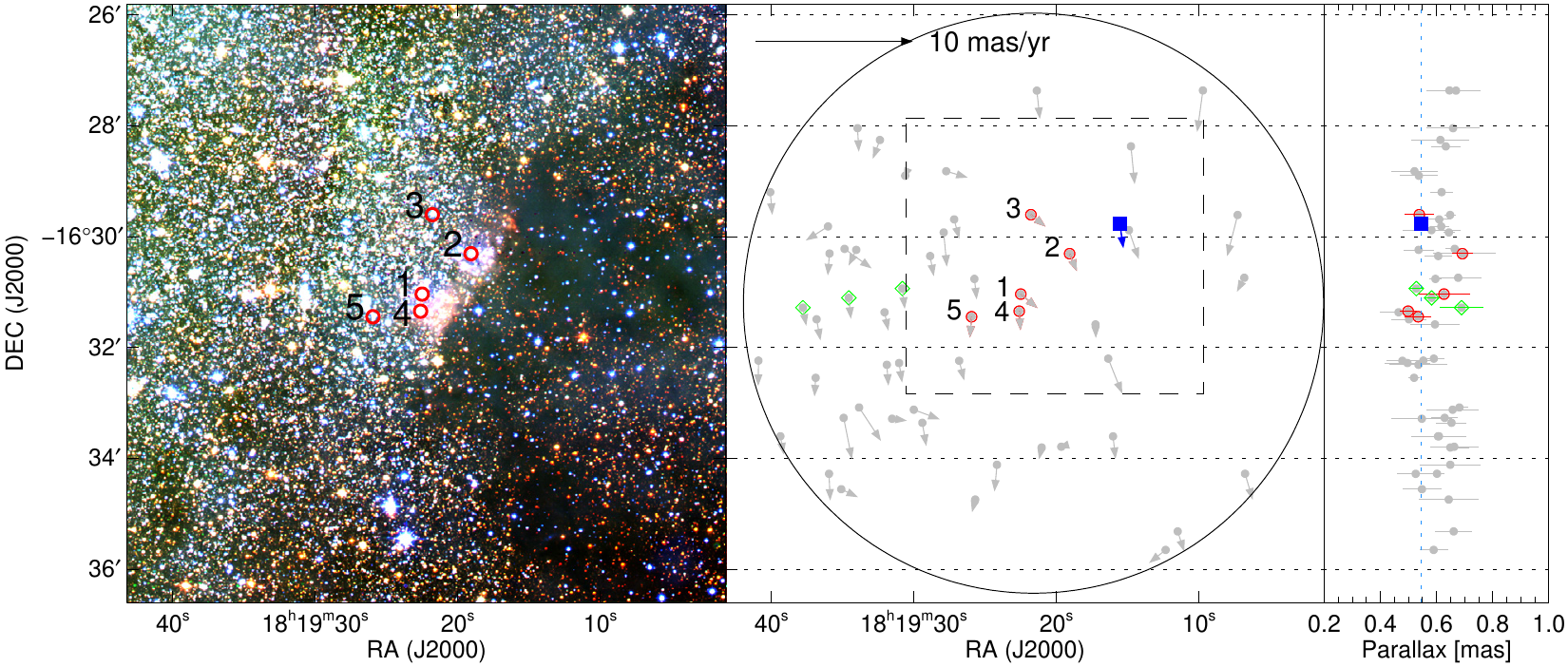}
      \caption{Candidate OBs selections around G014.645--00.606. A color-composite image presents the photometric observations of G014.645--00.606 from UKIDSS (blue: $J$, green: $H$, and red: $K_{\rm s}$). The red circles and black IDs indicate the five OB candidates (brightest stars) in the $K$ band. In the middle panel, the big circle represents the area to select OB candidates, and the dashed box indicates the zoom-in area of the right panel in Figure \ref{fig:m17}. The green diamonds indicate the three OB candidates far from IR bubbles. The blue square indicates the 22\,GHz water maser, which has a trigonometric parallax of $0.546\pm0.022$\,mas, measured by BeSSeL Survey of the VLBA \citep{2014A&A...566A..17W}. The gray points within a 5\,$\arcmin$ radius circle indicate the stars selected from Gaia EDR3, and the arrows indicate the proper motions. The parallaxes and errors of the stars are represented in the right panel.}\label{fig:imgjhk}
\end{figure}

\subsection{Selection of OB Candidates}
\label{sec:sel}

\begin{figure}[!ht]
 \centering
      \includegraphics[width=.6\textwidth]{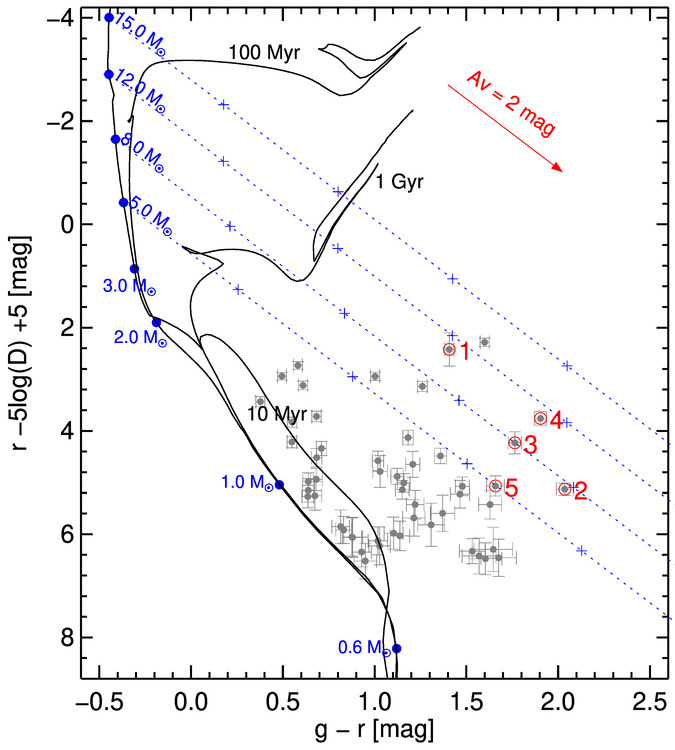}
      \caption{The CMD for the selected stars. The gray points with errors denote the selected stars from Gaia\,EDR3 in the circle region of Figure \ref{fig:imgjhk}. The potential OB stars are marked by red circles and IDs. The black curves present isochrones of 10\,Myr, 100\,Myr, and 1\,Gyr, and the blue points indicate the stellar mass in MS from 0.1\,$M_{\sun}$ to 15.0\,$M_{\sun}$, obtained from CMD\,3.5 \citep{2017ApJ...835...77M,2019MNRAS.485.5666P}. The red arrow denotes the extinction vector $A_V$. The distances, $D$, of the stars are obtained by the corresponding parallaxes in Gaia\,EDR3.}\label{fig:cmd}
\end{figure}

The left panel of Figure~\ref{fig:imgjhk} shows the $JHK$ color-composite image of G014.645--00.606. The nebula only seen in the $K$ band image is consistent with the $8.0\,\mu$m-bright rim. The five brightest stars in the $K$ band are close to the IR bubbles, as the red circles shown in the three panels. The parallax and the proper motions in R.A. and decl. of the EGO $G14.64-00.57$ were measured as $0.546\pm0.022$\,mas and ($0.22\pm1.20$, $-2.07\pm1.20$)\,mas yr$^{-1}$ \citep{2014A&A...566A..17W}. Because of the close relation between the EGO $G14.64-00.57$ and the IR bubbles, we suggest that the massive star candidates have parallaxes similar to that of the EGO $G14.64-00.57$. We have made use of the Gaia EDR3 data to identify the candidates, and the parallaxes of stars within a radius of $5\arcmin$ are constrained by conditions $\varpi/\sigma(\varpi)\geqslant 5$ and $0.4\arcsec\leqslant\varpi\leqslant 0.7\arcsec$. Some stars with similar proper motions may form a cluster, and the proper motions are concentrated in (0.2, -1.7)\,mas yr$^{-1}$ with a radius less than 1.0\,mas yr$^{-1}$. The cluster may be associated with EGO $G14.64-00.57$, and they have similar proper motion. Thus, the stars with the parallax constraints are considered when the difference between stellar proper motions and the proper motion (0.2, -1.7)\,mas yr$^{-1}$ is less than 2.0\,mas yr$^{-1}$. A total of 52 stars are close to EGO $G14.64-00.57$ in the parallax and proper motion. The gray points in the middle and the right panel of Figure~\ref{fig:imgjhk} show the proper motions and the parallaxes of the selected stars from Gaia EDR3, respectively. Then we introduce the second criterion that the absolute magnitude of massive star candidates should be analogous to the bona fide OB stars. 

We have used the photometric results of ZTF in the $g_\mathrm{ZTF}$ and $r_\mathrm{ZTF}$ bands to estimate the absolute  magnitude. Figure~\ref{fig:cmd} is the color-magnitude diagram (CMD) with the $g-r$ color as the horizontal axis and the $r - 5\log D + 5$ value as the vertical axis, where $D$ is the distance computed from the Gaia EDR3 parallax. The magnitudes and colors here have been converted from ZTF photometric system to Pan-STARRS1 (PS1; \citealt{2016arXiv161205560C}) photometric system, as shown in Appendix \ref{app:conv}, to match the photometric system in CMD\,3.5 \citep{2017ApJ...835...77M,2019MNRAS.485.5666P}. The distributions of the sources in Figure~\ref{fig:cmd} show two groups, the low-mass group lying close to the main-sequence locus in the mass range $0.6-1.5\,M_\odot$, and the higher-mass group lying in the lower right region where the extinction is also higher. Eight sources are lying within the mass range $3-12 \,M_\odot$, including the five luminous stars close to the IR bubbles. The other three stars that are far from the bubbles may not be relevant to the bubbles. 

\begin{table}
\bc
\begin{minipage}[]{100mm}
\caption[]{OB candidates \label{tab:obs}}\end{minipage}
\setlength{\tabcolsep}{1pt}
\footnotesize
 \begin{tabular}{cccccccccccc}
  \hline\noalign{\smallskip}
Source &  R.A. & Decl. & Parallax & $\mu_{\alpha}$ & $\mu_{\delta}$ &  $g_{\rm ZTF}$ & $r_{\rm ZTF}$ & Mass & log($T_{\rm eff}$) & $A_V$ &Binary  \\
 &  (J2000.0) & (J2000.0) & (mas) &  (mas yr$^{-1}$) &  (mas yr$^{-1}$) & (mag) & 
(mag) & ($M_{\sun}$)   & (K) & (mag) & \\
  \hline\noalign{\smallskip}
1 & 18:19:22.5 & -16:31:02 & 0.63(0.09) & 1.06(0.11) & -1.16(0.08) & 14.69(0.02)$^{\rm a}$ & 13.12(0.02)$^{\rm a}$ & 11.5(0.1)  & 4.42(0.01) & 5.8(0.1) & Y  \\
2 & 18:19:19.1 & -16:30:18 & 0.69(0.04) & 0.46(0.05) & -1.46(0.03) & 17.78(0.04)$^{\rm a}$ & 15.51(0.02)$^{\rm a}$ & 7.7(0.1) & 4.34(0.01) & 7.7(0.2) & Y  \\
3 & 18:19:21.8 & -16:29:36 & 0.54(0.05) & 0.93(0.07) & -0.99(0.05) & 17.27(0.03) & 15.27(0.02) & 8.2(0.1)   & 4.35(0.01) & 7.2(0.1) & -- \\                    
4 & 18:19:22.6 & -16:31:21 & 0.50(0.03) & 0.08(0.04) & -1.59(0.03) & 17.05(0.03) & 14.89(0.02) & 11.5(0.1) & 4.42(0.01) & 7.4(0.1) & -- \\                    
5 & 18:19:25.9 & -16:31:27 & 0.53(0.05) &-0.11(0.06) & -1.74(0.06) & 18.01(0.05) & 16.12(0.02) & 5.0(0.1) & 4.23(0.01) & 6.4(0.2) & -- \\    
  \noalign{\smallskip}\hline
\end{tabular}
\ec
\tablecomments{0.96\textwidth}{$^{\rm a}$ Photometry magnitudes of binaries are derived from the out-of-eclipse regions of the light curves. Parallax and proper motions ($\mu_{\alpha}$ and $\mu_{\delta}$) from the Gaia EDR3. The mass, temperatures, and extinction of the OB candidates are estimated from ZTF photometry.}
\end{table}

Since the position of a star in CMD is constrained by the mass and age of the star, we have estimated the masses of the five stars on different isochrones along the extinction vector $A_V$ in Figure~\ref{fig:cmd}. The visual extinction $A_V$ is treated as a free parameter. The convention between $A_V$ and the extinction in the $g$ and $r$ bands is $A_g/A_V=1.155$ and $A_r/A_V=0.843$ \citep{2019ApJ...877..116W}, respectively. Assuming a value of $A_V$, the extinction $A_r$ in the $r$ band is then computed according to the corresponding ratio. The stellar mass inferred from the CMD depends strongly on the assumed age. Based on the isochrones of 1\,Gyr, 100\,Myr, and 10\,Myr in Figure 3, the masses of the five stars are estimated to be in the range of about 2\,M$_{\sun}$, 4-5\,M$_{\sun}$, and 5-12\,M$_{\sun}$, respectively. At the assumed age of 10 Myr, the five stars fall into the types of early-B stars. We have estimated the stellar parameters, such as, mass, temperature, and extinction of the five stars according to the 10\,Myr isochrone, as listed in Table \ref{tab:obs}.   

Sources 2, 4 and 5 are OB candidates earlier than B3V classified by \citet{2019MNRAS.487.1400C}. The three candidates are identified from the VPHAS+ data via the color-color diagram. The different results between this work and \citet{2019MNRAS.487.1400C} may be due to the different photometric data and selection method of OB candidates. In addition, we considered the distance in the CMD, and sources 1 and 3 are then classified as OB candidates.

\subsection{The Most Probable OB Stars Associated with the Radio-quiet \ion{H}{ii} region G014.645-00.606 }
\label{sec:property}
Because there are obvious H$\alpha$ emissions and mid-IR radiation around the sources 1, 2, and 3 in Figure \ref{fig:m17}, they may be associated with the corresponding bubbles and cause H$\alpha$ emissions and mid-IR radiation to form the bright rim. Source 1 is wrapped by the brightest H$\alpha$ emissions in bubble-1 with a radius of 0.35\,pc, and has a closer distance of about 0.12\,pc to the mid-IR rim. The H$\alpha$ emissions extend in a gradient weakening to the northeast, which suggest that source 1 is one of OB stars excited the \ion{H}{ii} region. Sources 2 and 3 are in bubbles of radius 0.22\,pc and 0.44\,pc respectively, and accompanied by the relatively weak H$\alpha$ emissions. However, source 4 lies on the boundary of bubble-1, and its relation to surrounding radiation is not clear. The ionization characteristics of source 4 may be obscured by those of source 1 due to the larger distance. Source 5 is outside the boundary of bubble-1 and may not be directly related to ionized region.

Given that an OB star drives \ion{H}{ii} region, the \ion{H}{ii} region expands rapidly to an equilibrium state as the Str{\"o}mgren sphere with a radius \citep{2007ApJ...671..518K} of 

\begin{equation}
{r_s = (\frac{3S_{\rm Ly}}{4\pi\alpha^{(\rm B)}n_{\rm H}^2})^{1/3}},
\end{equation}
where $\alpha^{(\rm B)}$ is the case B recombination coefficient, $\alpha^{(\rm B)}\approx2.59\times10^{-13}(T/10^4 {\rm K})^{-0.7}$\,cm$^{-3}$s$^{-1}$ \citep{1989agna.book.....O} at a gas temperature $T$. The $n_{\rm H}$ is the density of neutral gas. Assuming a typical value of $n_{\rm H}=100\,\mathrm{cm^{-3}}$ in Galactic molecular clouds, the ionized gas temperature of 8,000\,K \citep{2007ApJ...671..518K}, and $N_\mathrm{LyC}\leqslant1.0\times10^{45}\,\mathrm{photons\,s^{-1}}$, the Str{\"o}mgren radius of \ion{H}{ii} region is computed as 0.1\,pc. An early-B star with a temperature of about 23,000\,K can drive the \ion{H}{ii} region with a radius of 0.1\,pc and a number of ionizing photons $1.0\times10^{45} \,\mathrm{photons\,s^{-1}}$. The radii of the bubbles around sources 1, 2, and 3 seem larger than the Str{\"o}mgren radius of early B-type stars, which could be the result of the expansion of the \ion{H}{ii} regions. And the radius increases with time and strongly depends on the surrounding. When the \ion{H}{ii} region expands, the temperature of the inner gas is higher than that of the surrounding cloud material, and the ionization front expands under the pressure inside the ionized gas \citep{2020MNRAS.492..915G}. From the positions and radii of the bubbles, we believe that sources 1, 2 and 3 could be the most early B-type stars driving the bubbles.


\section{Early-B Eclipsing Binaries}
\label{sec:obvar}
We have compared the early-B candidates (sources 1, 2, and 3) with the public catalog of variable stars, and found that source 1 is a known variable star V1963\,Sgr in the General Catalog of Variable Stars (GCVS; \citealt{2017ARep...61...80S}), International Variable Star Index (VSX; \citealt{2006SASS...25...47W}), and All-Sky Automated Survey for Supernovae (ASAS-SN; \citealt{2018MNRAS.477.3145J}), as listed in Table \ref{tab:ob1}. Source 1 is an eclipsing binary of Algol type (EA) classified by GCVS and VSX, while it seems an eclipsing binary of W Ursae Majoris type (EW) from the light curves of ASAS-SN. We have analyzed the variabilities of the binaries in this section, including period and variability type, color variability, physical parameters, and long-term variation of the period.

\begin{table}
\bc
\begin{minipage}[]{100mm}
\caption[]{The observations of sources 1 and 2\label{tab:ob1}}\end{minipage}
\setlength{\tabcolsep}{1pt}
\small
 \begin{tabular}{llrrccc}
  \hline\noalign{\smallskip}
Observation &  Time interval & Bands & Number of obs. & Period (days)& Variability type & Ref.  \\
  \hline\noalign{\smallskip}
GCVS 		& 2436755.5--2441068.6 & PE$^a$ 	& 233 & 1.407205 & EA & 1,2 \\
VSX 		& 2459312.9--2459381.9 & CV$^b$ 	& 7     & 0.825075 & EA & 3 \\
ASAS-SN 	& 2457078.1--2458376.5 & $V$ 		& 395 & 0.825070 & EW & 4 \\
ZTF 		& 2458245.0--2459391.8 & $g_{\rm ZTF}$ and $r_{\rm ZTF}$& 91, 206 & 0.825079$^c$ & EW & 5,6 \\
YAHPT 	& 2459403.1--2459434.1 & $V$, $R_c$, and $I_c$ & 191, 262, 258 & 0.825319 & EW & 6\\
\hline
\multicolumn{7}{c}{source 2}\\
\hline
ZTF 		& 2458245.0--2459391.8 & $g_{\rm ZTF}$ and $r_{\rm ZTF}$& 91, 206 & 0.919379$^c$ & EW & 5,6 \\
YAHPT 	& 2459403.1--2459434.1 & $V$, $R_c$, and $I_c$ & 126, 262, 257 & 0.919883 & EW & 6\\
  \noalign{\smallskip}\hline
\end{tabular}
\ec
\tablecomments{0.86\textwidth}{$^a$Photographic emulsion. $^b$Clear (unfiltered) reduced to $V$ sequence. $^c$Period is calculated in this work.}
\tablerefs{0.86\textwidth}{(1) \citet{1991IBVS.3574....1M}, (2) \citet{2017ARep...61...80S}, (3) \citet{2006SASS...25...47W}, (4) \citet{2018MNRAS.477.3145J}, (5) \citet{2019PASP..131a8003M}, (6) This work.}
\end{table}

\subsection{Period and Variability Type}
\label{sec:var}
\begin{figure}[!ht]
 \centering
          \includegraphics[width=.99\textwidth]{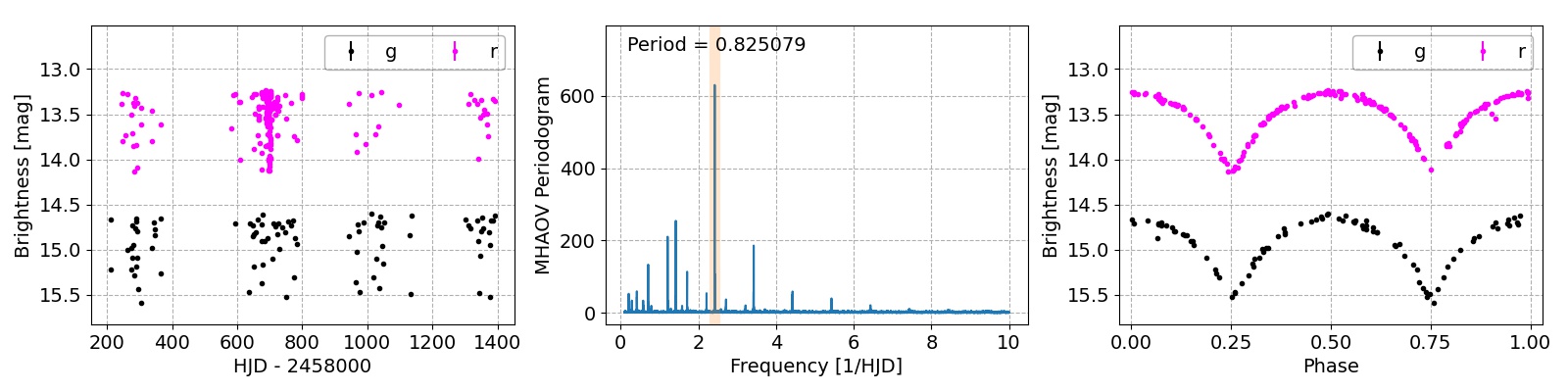}
          \includegraphics[width=.99\textwidth]{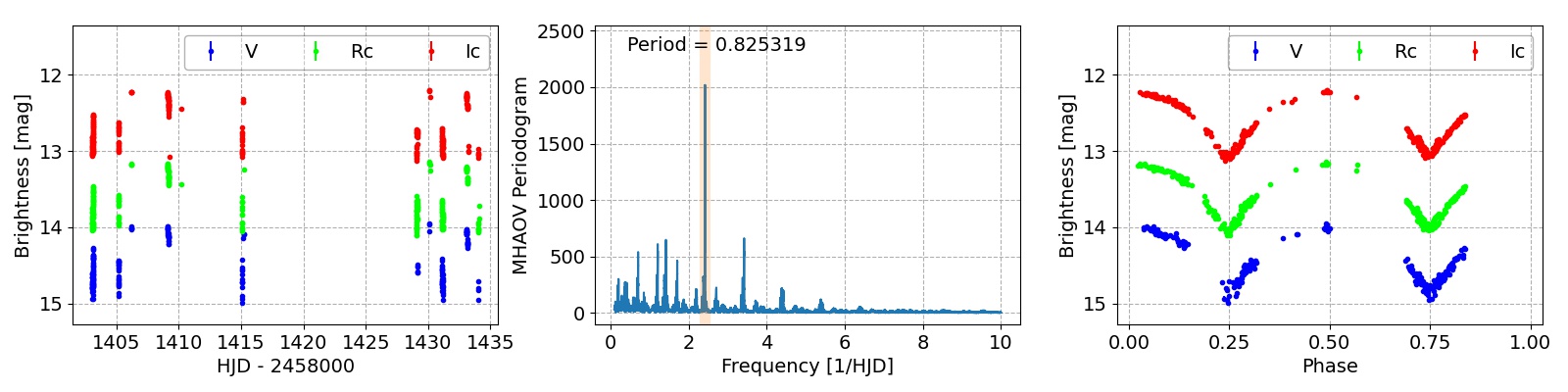}
      \caption{Light curves, periodograms, and folded light curves of source 1. The first and second rows are the periodic analysis results from ZTF and YAHPT observations, respectively. The typical photometric errors are less than 0.02\,mag in $g_{\rm ZTF}$, $r_{\rm ZTF}$ band, $V$,  $R_c$, and $I_c$ bands.}\label{fig:aov1}
\end{figure}

\begin{figure}[!ht]
 \centering
          \includegraphics[width=.99\textwidth]{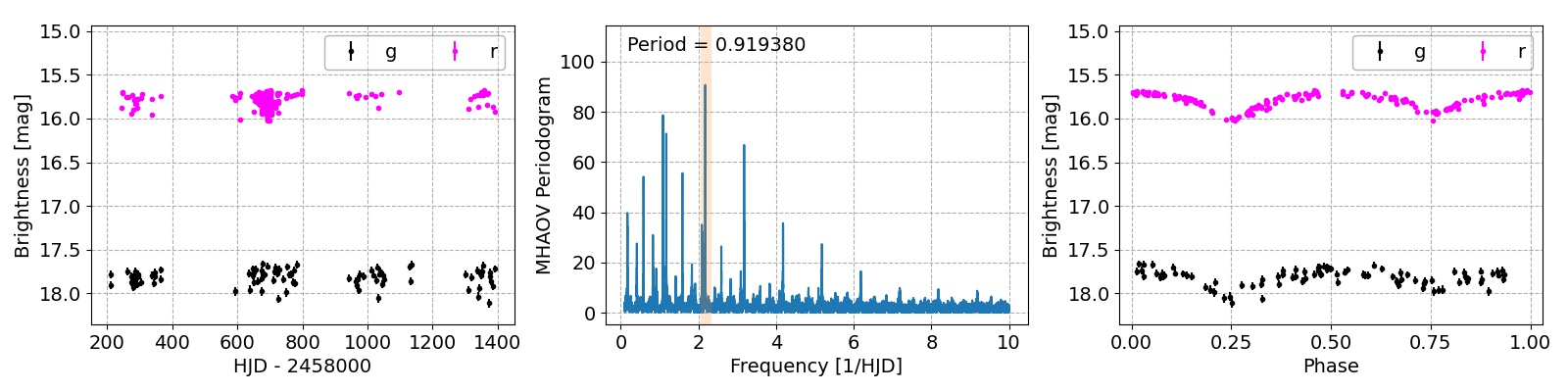}
          \includegraphics[width=.99\textwidth]{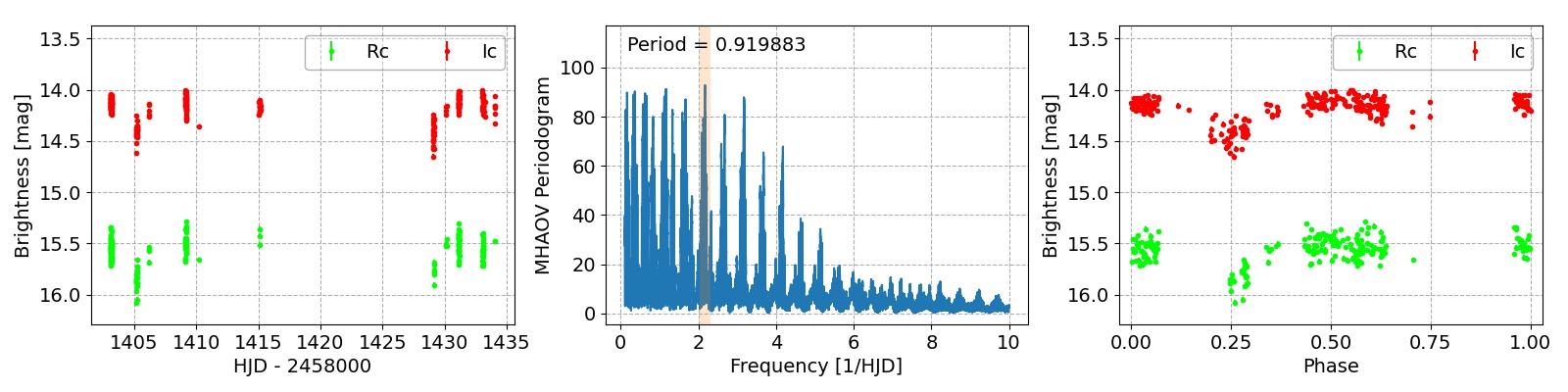}
      \caption{Light curves, periodograms, and folded light curves of source 2. The typical photometric errors are less than 0.02\,mag in $r_{\rm ZTF}$ band, less than 0.03\,mag in $R_c$ and $I_c$ bands, and less than 0.05\,mag in $g_{\rm ZTF}$ band.}\label{fig:aov2}
\end{figure}

In order to research the variability of the three early-B candidates, we have used the time-series photometric observations of ZTF, which covered from May 2018 to June 2021, and about 100 observations in $g_{\rm ZTF}$ band and 200 observations in $r_{\rm ZTF}$ band. The mutilband Analysis of Variance (AoV; \citealt{2015ApJ...811L..34M}) method is used to analyze observations, and the sampling frequency is set to 0.1--10 days$^{-1}$ under a frequency spacing of $\Delta f = 1.0\times 10^{-5}$. The folded light curves of sources 1 and 2 show significant amplitude and periodic variations, while those of source 3 show no obvious amplitude and periodic variation. The amplitudes of the light curves are about 0.9\,mag and 0.3\,mag for sources 1 and 2, respectively. The periodic analysis of sources 1 and 2 are shown in Figure \ref{fig:aov1} and Figure \ref{fig:aov2}, respectively. Due to the coincidence of the primary and secondary eclipses of binary stars, the half periods of the binary stars can be calculated from the optimal frequency of the AoV method, and the period of sources 1 and 2 is 0.82508\,day and 0.91938\,day, respectively. The typical photometric errors of source 1 are less than 0.02\,mag in the two bands, while the typical photometric errors of source 2 are 0.05\,mag in $g_{\rm ZTF}$ and 0.02\,mag in $r_{\rm ZTF}$ bands.

We have used the high-cadence observation of YAHPT in $V$, $R_c$, and $I_c$ bands (as listed in Table \ref{tab:bands}) to verify the periods of sources 1 and 2. The typical photometric errors of source 1 are less than 0.02\,mag in the three bands. For source 2, the photometric data in $V$ band is not used due to unavailable photometric quality, the photometric data in $R_c$ and $I_c$ bands with errors less than 0.03\,mag have been used. The same periodic analysis is applied to the three bands, which have about 300 observations in each band. The periods of sources 1 and 2 are 0.82532\,day and 0.91988\,day, which are slightly different from the previous results and suggest a long-term variation of the periods (in Section \ref{sec:oc}). 

As shown in Figure \ref{fig:aov1}, the folded light curves indicate an EW-type binary system of source 1, and the same primary and secondary eclipses imply that the two components have nearly the same stellar temperatures. In Figure \ref{fig:aov2}, source 2 is a newly discovered binary system, whose primary and secondary eclipses are different in amplitude, and the temperatures of the components are definitely different. As the short-period B-type eclipsing binaries (EB), they are important due to rarely found so far. In a new catalog of OBA-type stars observed by TESS, there are 3425 EBs and 737 B-type EBs, and only 57 B-type EBs have a period of less than 1.0\,days \citep{2021A&A...652A.120I}.

\subsection{Color Variability}
\label{sec:color}
\begin{figure}[!ht]
 \centering
          \includegraphics[width=.95\textwidth]{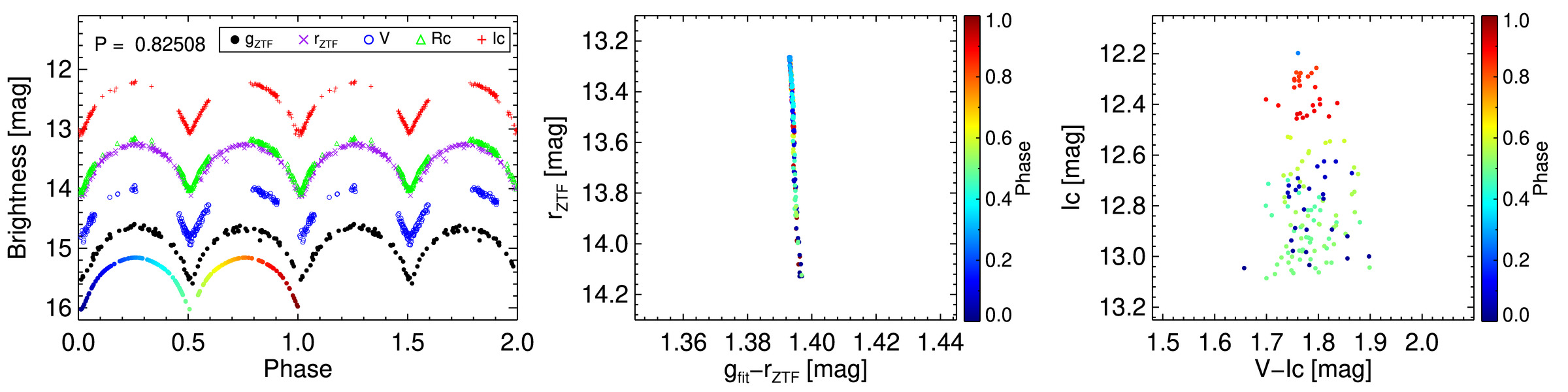}
          \includegraphics[width=.95\textwidth]{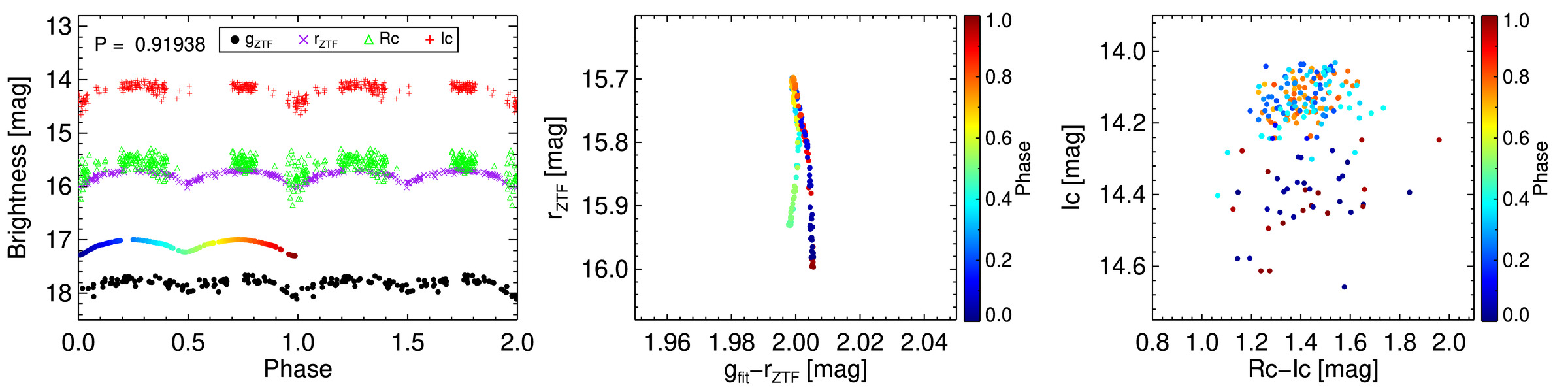}
      \caption{Color variability of sources 1 and 2. The first column shows the folded light curves in $g_{\rm ZTF}$, $r_{\rm ZTF}$, $V$, $R_c$, and $I_c$, and the fitting results ($g_{\rm fit}+constant$) of $g_{\rm ZTF}$ are indicated by color, $g_{\rm fit}+0.5$ for source 1 and $g_{\rm fit}-0.7$ for source 2. The second column shows the CMDs constructed from $g_{\rm fit}$ and $r_{\rm ZTF}$, where phase is indicated by color. The third column shows the CMDs constructed from $V$, $R_c$ and $I_c$ in YAHPT observations, and the phase is indicated by color. The typical error is less than 0.02\,mag for $g_{\rm fit}-r_{\rm ZTF}$ and 0.06\,mag for $R_c-I_c$, respectively.}\label{fig:cmds}
\end{figure}

Color variability of eclipsing binary shows a particular pattern in the CMD, as described by \citet{2019A&A...623A.110G}. The variability of eclipsing binary systems induces linear motions in a time-dependent CMD, and the motions are vertical with great amplitudes in magnitude and small amplitudes in color, as shown in Figure \ref{fig:cmds}. 

We have used ZTF and YAHPT observations to produce the color variability of eclipsing binaries in CMDs respectively. Due to non-simultaneous observations between $g_{\rm ZTF}$ and $r_{\rm ZTF}$, a model result of the eclipsing binary systems in $g_{\rm ZTF}$ band in Section \ref{sec:para} is used to compute the color $g_{\rm ZTF}-r_{\rm ZTF}$. For the colors of YAHPT observation, $V -I_c$ and $R_c - I_c$, the two pair observations would be selected when their time interval is less than 0.001\,day, which is far less than the limitation of the pattern variations caused by non-simultaneous observation \citep{2021AJ....162...52Y}.

As expected, the vertical motions of sources 1 and 2 induced by color variability in CMD are caused by the ecliping of the binaries. Because stellar rotation could cause the motions with a smaller amplitude in magnitude, and the extinction could cause the motions in the direction of the extinction vector \citep{2019A&A...623A.110G}, the motions of sources 1 and 2 are not caused by the rotation and extinction of the star. Due to the smaller error in $r_{\rm ZTF}$ band and the introduction of the fitting result in $g_{\rm ZTF}$ band, the motions of the ZTF observations are more convergence than those of YAHPT observations. The nearly invariable colors of source 1 suggested the two components have almost the same temperature, while the apparent dispersion in the color variability of source 2 may be ascribed to the eclipsing of the two components with different temperatures.  

According to the Table\,5 of \citet{2013ApJS..208....9P}, we have estimated the SpTs of the binaries sources 1 and 2 from the effective temperatures of Table \ref{tab:obs}. The SpTs of the binaries sources 1 and 2 are B1V and B1.5V, respectively. However, the SpT of source 1 was classified as F8/G0e + F/G$?$ by \citet{1984PASP...96...98H}, and the spectrum appears veiled and composite with a H$\beta$ emission peak and some double emissions. The spectral divergence may be caused by the envelope extinction of of the contact binary system, and the spectrum was veiled by the envelope of the contact binary source 1. For the components of binary systems, the brightness of the primary star will be lower. We have estimated the primary temperature ($T_1$) and the secondary temperature ($T_2$) in CMD, based on the estimated temperature ratios ($T_2/T_1$) of the light curves. The temperature ratios of sources 1 and 2 are about 1.0 and 0.8, and the primary temperatures of sources 1 and 2 are estimated to be 23,513\,K and 20,145\,K, equivalent to SpTs of B2V and B2V, respectively.

\subsection{Modeling of Eclipsing Binaries}
\label{sec:para}
We have used the PHysics Of Eclipsing BinariEs (PHOEBE; \citealt{2005ApJ...628..426P}) to derive the parameters of the eclipsing binaries found with the ZTF observations. The PHOEBE, based on the Wilson--Devinney code \citep{1971ApJ...166..605W} is often used to reproduce and fit the light curves, radial velocity curves, and spectral line profiles of eclipsing systems though modeling eclipse events \citep{2016ApJS..227...29P}. 

\begin{table}
\bc
\begin{minipage}[]{100mm}
\caption[]{The parameters of the modeled binary systems\label{tab:phoebe}}\end{minipage}
\setlength{\tabcolsep}{1pt}
\small
 \begin{tabular}{lcc}
  \hline\noalign{\smallskip}
OBs\_ID                       & Source 1       & Source 2      \\
Classification             & EW                 & EA  \\
  \hline\noalign{\smallskip}
\multicolumn{3}{c}{Measured Parameters}\\
\hline
Amplitude of $r_{\rm ZTF}$ (mag) 	& 0.887                         & 0.338         \\
Period $P$ (day) 			& 0.82508                  & 0.91935     \\
\hline
\multicolumn{3}{c}{Free Parameters}\\
\hline
CMD Temperature $T_1$ (K) & 23513 (fixed) 		& 20145 (fixed)  	 \\
Temperature ratio $T_2/T_1$& 0.993$_{-0.009}^{+0.009} $   & 0.772$_{-0.024}^{+0.019} $     \\
Epoch $t_0$ (MJD-58000)$^a$ & 0.4609$_{-0.0004}^{+0.0004} $ & 0.8708$_{-0.0017}^{+0.0015} $  \\
Mass ratio $q$ & 0.98$_{-0.16}^{+0.15} $           & 0.73$_{-0.17}^{+0.3} $   \\
Inclination $i$ (deg)& 86.08$_{-0.43}^{+0.61} $          & 67.7$_{-1.1}^{+1.1} $  \\
Semi-major axis $a$ ($R_{\odot}$) & 11.5$_{-3.0}^{+6.0} $        & 12.7$_{-2.1}^{+5.0} $  \\
Equivalent Radius $R_1$ ($R_{\odot}$) & 4.8$_{-1.2}^{+2.5} $  & 5.18$^b$   \\
Equivalent Radius $R_2$ ($R_{\odot}$) & 4.76$^b$  &  3.33$_{-0.59}^{+1.17}$\\
Potentials$^b$ $\Omega$ & 3.525                    & -           \\
Fill-out factor $f^b$ & 0.361                    & -     \\
pbLum$_r^c$ & 5.79$_{-0.37}^{+0.48} $            & 9.28$_{-0.54}^{+0.56} $     \\   
  \noalign{\smallskip}\hline
\end{tabular}
\ec
\tablecomments{0.86\textwidth}{$^{\rm a}$ The phase value of the primary eclipse is given here rather than the time. $^{\rm b}$ The constraint parameters of the PHOEBE model. The value of $R_1$ (or $R_2$) constrains $\Omega$, $f$, and $R_2$ (or $R_1$). The fillout factor of the system is defined as $(\Omega-\Omega_{L_1})/(\Omega_{L_2}-\Omega_{L_1})$, $L_1$ and $L_2$ represent \emph{Lagrangian points}, and $L_1$ is the contact point of the Roche lobe \citep{2018maeb.book.....P}. $^{\rm c}$ The best-fitting results of bandpass luminosity for the normalized fluxes in $r$.}
\end{table}

PHOEBE provides default values for the parameters of the binary systems, and contains free parameters and read-only parameters due to built-in constraints \citep{2020ApJS..250...34C}. In this work, the built-in constraints have been preserved to keep the model within the real of physical solutions, and the values of some free parameters are obtained from the observations, as shown in Table~\ref{tab:phoebe}. The periods ($P$), effective temperature of each primary star ($T_1$), and extinction ($A_V$) values obtained from previous analysis are used. The $A_V$ is used to adjust the observed luminosity ratio to the intrinsic luminosity ratio between $g_{\rm ZTF}$ and $r_{\rm ZTF}$ bands. The fluxes in the $g_{\rm ZTF}$ and $r_{\rm ZTF}$ bands are used as the `dataset-coupled' mode of the PHOEBE model, which means the same scaling factor of passband luminosity to be applied to two different datasets, and the $g_{\rm ZTF}$ fluxes is bound to the $r_{\rm ZTF}$ fluxes. Because the CMD temperatures of the components are all above 6000\,K, the bolometric gravity brightening, $g = 1.0$, and albedo, $A = 1.0$, values suggested by PHOEBE are used when modeling the light curves. The eccentricity (Ecc) of the contact binaries source 1 and semi-detached binary source 2 is set to a default value of zero.

\begin{figure}[!ht]
 \centering
          \includegraphics[width=.47\textwidth]{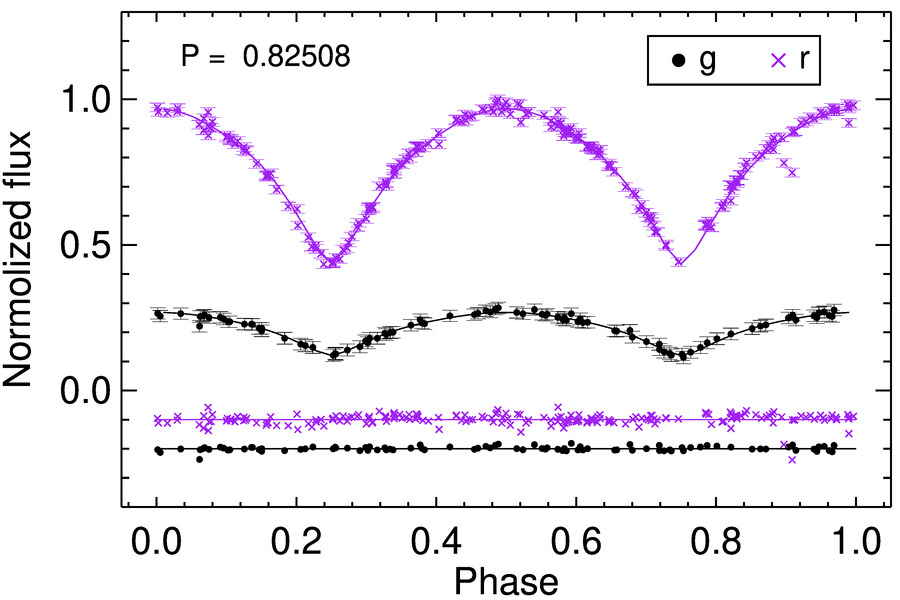}
          $\quad$
          \includegraphics[width=.47\textwidth]{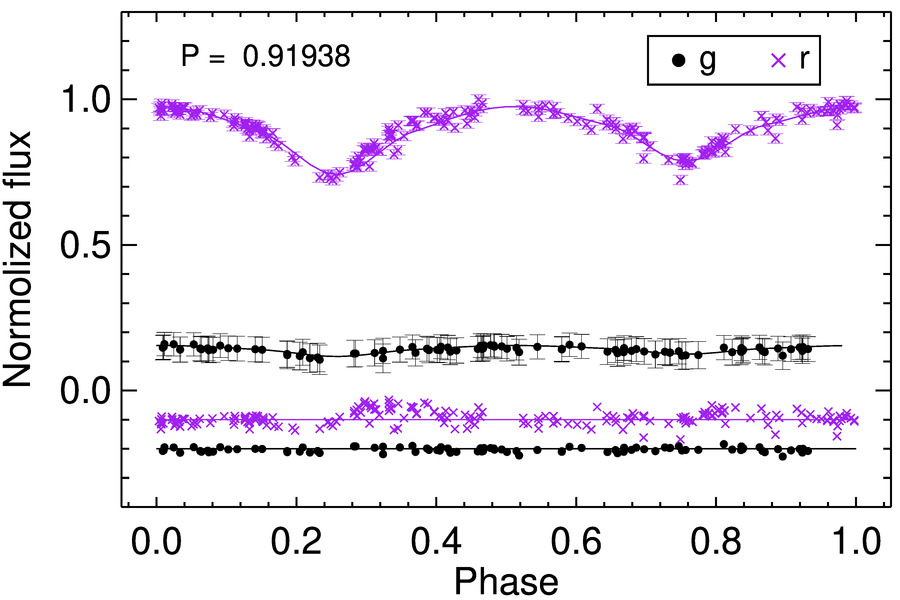}
      \caption{Light-curve modeling results of sources 1 and 2. The black points and purple crosses represent the observations in $g_{\rm ZTF}$ and $r_{\rm ZTF}$. The curves denote the corresponding fitting results. The residuals ($\delta g_{\rm ZTF}-0.2$ and $\delta r_{\rm ZTF}-0.1$) are shown at the bottom of the figures.}\label{fig:fitting}
\end{figure}

After optimizing and determining a reasonable starting model by hand, the fitting models to the light curves of sources 1 and 2 are performed via an affine-invariant Markov chain Monte Carlo (MCMC) ensemble sampling code EMCEE \citep{2013PASP..125..306F}. We have used the uniform priors for the free parameters in Table~\ref{tab:phoebe}, and set 1200 steps and 20 walkers for sampling the parameter space to explore the posterior distributions of the free parameters. Due to the different variability types of sources 1 and 2, the models of sources 1 and 2 are configured as contact binary system and semi-detached binary system respectively. The optimized parameter estimates and the uncertainties of posterior density are listed in Table~\ref{tab:phoebe}, and the posterior distributions are illustrated in Appendix \ref{app:emcee}. The optimized light curve models constructed from these parameters are shown in Figure \ref{fig:fitting}. Due to the mass ratio of 0.98, the temperature ratio of 0.993, and about 4.8\,$R_{\odot}$ radii of primary and secondary stars, source 1 could be a twin binary. From the fill-out factor of envelope 0.361, the twin binary system has contact with a common envelope, and the mass exchange may have occurred. 

\subsection{Possible Decrease of Orbiting Period of Source 1}
\label{sec:oc}
\begin{figure}[!ht]
 \centering
          \includegraphics[width=.52\textwidth]{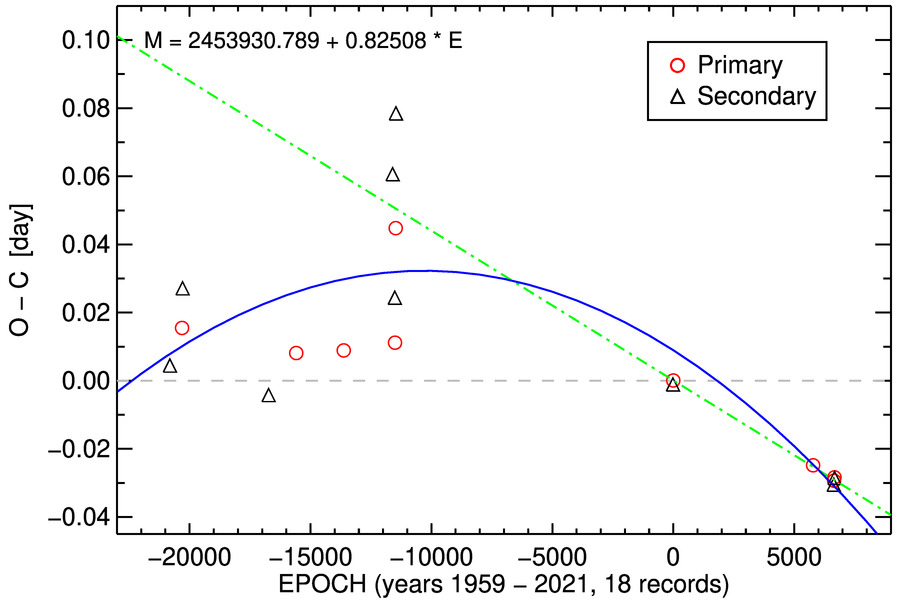}
      \caption{The $O-C$ diagram for the minimum time of source 1. The last five points are calculated from the ZTF and YAHPT observations, and the points in front are from \citet{1991IBVS.3574....1M} and \citet{2007IBVS.5806....1K}. The blue line shows the parabolic fitting result of $O-C$ values. The green dot-and-dash line shows the line fitting result of the last seven $O-C$ values.}\label{fig:oc}
\end{figure}

The temporal variation in the period of EB provides useful information about mass transfer, presence of third body, and other characteristics. We have used the $O-C$ Diagram to study the temporal variation of EB period, some of the minimum times ($O$) of source 1 are obtained from the Eclipsing Binaries Minima Database\footnote{\url{http://var2.astro.cz/ocgate/}} \citep{2006OEJV...23...13P}. Based on the observations of $r_{\rm ZTF}$, $V$, $Rc$, and $Ic$ bands in Table \ref{tab:bands}, the minima and uncertainties of sources 1 and 2 are calculated by using Nelson's polynomial fitting program\footnote{Software by Bob Nelson, \url{http://binaries.boulder.swri.edu/binaries/software/}}, which was developed by the method of \citet{1956BAN....12..327K}. As shown in Table \ref{tab:min} and Figure \ref{fig:oc}, six new minima are derived, including five minimum times (three primary, P-type, and two secondary, S-type, minima) for source 1 and one P-type minimum time for source 2. The calculated times can be expressed as $C = T_0 + P \times E$, where $E$ is the epoch number represented as the integer value of $(O-T_0)/P$ and its decimal number corresponds to the $O-C$ value. As the recent period results of source 1 are about 0.825\,day, we adopt the previous maximum epoch of $T_0 = 2453930.789$ and the period of $P = 0.82508$\,days in \citet{2007IBVS.5806....1K}. The ephemerides of sources 1 and 2 are as follows:
$$Min.I(HJD) = 2453930.7890(\pm 0.0002) + 0.82508(\pm 0.00001)E$$
$$Min.I(HJD) = 2458691.7502(\pm 0.0007) + 0.91938(\pm 0.00002)E$$

\begin{table}
\bc
\begin{minipage}[]{100mm}
\caption[]{Photometric data (the first five rows) of sources 1 and 2 in different bands\label{tab:bands}}\end{minipage}
\setlength{\tabcolsep}{1pt}
\footnotesize
 \begin{tabular}{cccccc|cccc}
  \hline\noalign{\smallskip}
HJD, $V$& $V$ & HJD, $R_c$& $R_c$ & HJD, $I_c$ & $I_c$  & HJD, $R_c$& $R_c$ & HJD, $I_c$ & $I_c$  \\
(245,+)  &(mag)  &(245,+) &(mag) &(245,+) & (mag)&(245,+) &(mag)  &(245,+) & (mag)\\
  \hline\noalign{\smallskip}
\multicolumn{6}{c}{source 1} & \multicolumn{4}{c}{source 2}\\
\hline
9403.073112 & 14.603(0.014) & 9403.073584 & 13.760(0.005) & 9403.074057 & 12.817(0.004) & 9403.073583 & 15.602(0.024) & 9403.074057 & 14.091(0.011) \\
9403.075148 & 14.649(0.014) & 9403.074602 & 13.746(0.005) & 9403.075634 & 12.857(0.004) & 9403.074602 & 15.345(0.019) & 9403.075633 & 14.190(0.013) \\
9403.076608 & 14.621(0.012) & 9403.077091 & 13.797(0.005) & 9403.076080 & 12.818(0.004) & 9403.077090 & 15.360(0.019) & 9403.076080 & 14.038(0.012) \\
9403.077627 & 14.598(0.012) & 9403.078667 & 13.807(0.005) & 9403.078114 & 12.817(0.004) & 9403.078667 & 15.343(0.020) & 9403.078113 & 14.114(0.011) \\
9403.080149 & 14.652(0.013) & 9403.079134 & 13.801(0.005) & 9403.079616 & 12.863(0.004) & 9403.079133 & 15.540(0.024) & 9403.079616 & 14.069(0.011) \\
  \noalign{\smallskip}\hline
\end{tabular}
\ec
\end{table}

\begin{table}
\bc
\begin{minipage}[]{100mm}
\caption[]{The new minimum times of sources 1 and 2 in different bands \label{tab:min}}\end{minipage}
\setlength{\tabcolsep}{1pt}
\small
 \begin{tabular}{lllllllc}
  \hline\noalign{\smallskip}
HJD, $r_{\rm ZTF}$   & HJD, $V$ & HJD, $R_c$& HJD, $I_c$ & Mean& Epoch &$O-C$  & Type      \\
(245,+)  &(245,+)  &(245,+)  &(245,+) &(245,+)  &  & (days) & \\
  \hline\noalign{\smallskip}
\multicolumn{8}{c}{source 1 (V1963 Sgr)}\\
\hline
8699.72570(7.6E-4)& & & & 8699.72570(7.6E-4) & 5780 & -0.025 & P \\                                          
 & 9403.10265(4.0E-4) & 9403.10346(1.5E-4) & 9403.10275(2.5E-4) & 9403.10295(2.7E-4)  & 6632 & -0.031 & S \\ 
 &    & 9415.06782(3.9E-4) & 9415.06840(8.0E-4) & 9415.06811(6.0E-4)  & 6647 & -0.029 & P \\ 
 &    & 9429.09240(4.9E-4) & 9429.09225(2.5E-4) & 9429.09232(3.7E-4)  & 6664 & -0.028 & P \\ 
 & 9431.15720(2.6E-4) & 9431.15435(3.5E-4) & 9431.15625(4.7E-4) & 9431.15593(3.6E-4) & 6666 & -0.029 & S \\ 
 \hline
\multicolumn{8}{c}{source 2}\\
\hline
8691.75020(7.0E-4)& & & & 8691.75020(7.0E-4) & 0  &0  &P \\ 
  \noalign{\smallskip}\hline
\end{tabular}
\ec
\end{table}

As shown in the Figure \ref{fig:oc}, the $O-C$ values of source 1 cover more than six decades, and show a period decrease in recent decades. The aperiodic behaviour does not support the third body case that causes the periodic variation, and may be attributed to mass transfer or mass loss from one component to the other \citep{2010JApA...31...97L}. The $O-C$ values of source 1 can be fitted by a downward parabolic curve, which can be represented by the following equation:
$$Min.I(HJD) = 2453930.798(\pm 0.007) + 0.82508(\pm 0.0002)E - 2.2(\pm 0.7)\times10^{-10}E^2$$
The quadratic term, $-2.2\times10^{-10}$, means that the orbital period decreases at a rate of $-2.0(\pm 0.6)\times10^{-7}$\,days yr$^{-1}$. Due to $q<1$ derived in Section \ref{sec:para}, the period change rate of source 1 suggests that the mass transfer is taking place from the primary to secondary component and causes the orbit to shrink \citep{2015IJAA....5..222N}.

Unfortunately, the photometric uncertainty in the study of \citet{1991IBVS.3574....1M} is about 0.25\,mag, the $O-C$ values may have a large deviation. Based on the $O-C$ values of \citet{2007IBVS.5806....1K} and this work, the orbital period decreases at a rate of $-1.90(\pm 0.01)\times10^{-3}$\,days yr$^{-1}$, as green dot-and-dash line shown in Figure \ref{fig:oc}. The negative slope points to a real period that is shorter than the used period \citep{2005ASPC..335....3S}. However, such a small amount of data is insufficient to account for the result, the variant pattern of orbital period needs to be confirmed by more observations.

\section{Conclusions}
\label{sec:con}
In this work, we have used the H$\alpha$ and IR images, astrometric information of Gaia EDR3, and ZTF photometric observations to identify OB candidates in the radio-quiet \ion{H}{ii} region G014.645--00.606 in M17 SWex, and found that three bright stars (sources 1, 2 and 3) are the most likely candidates related to the \ion{H}{ii} region and located in the mid-IR bubbles. The three candidates are identified as early B-type stars, which can drive the observed IR bubbles with a radius of about 0.2\,pc and a number of ionized photons of about $10^{45} \,\mathrm{photons\,s^{-1}}$.

We have made use of the ZTF and YAHPT time-series photometric observations to analyze the variabilities of the candidates. Based on the AoV method, source 1 (V1963 Sgr) and source 2 are proven to be EBs, source 1 is an EW type star with a period of 0.82508\,day, and source 2 is an EA type star with a period of 0.91935\,day. From the brightness of the two binary systems in CMD, the primary temperature of sources 1 and 2 is about 23,000\,K and 20,000\,K respectively. The physical parameters have been fitted by PHOEBE model from light curves and showed that source 1 could be a twin binary. The new minimum times of sources 1 and 2 are added, and the period of source 1 show long-term variations in the $O-C$ diagram. The orbital period of source 1 decreases at a rate of $-2.0(\pm 0.6)\times10^{-7}$\,days yr$^{-1}$, suggesting mass exchange between the components. However, the number of the minimum times in the $O-C$ diagram is still small, the variant pattern of orbital period needs more observation to be confirmed.

The massive binary systems associated with \ion{H}{ii} regions provide us with more opportunities to study the evolution of massive stars, such as single star evolution or binary mergers. Our next work will be to establish the multiplicity properties of large samples of OB stars from the \ion{H}{ii} regions and statistical analyze the physical parameters to constraint the evolution mechanism of massive binary systems.

\normalem
\begin{acknowledgements}
This work is supported by National Key Research \& Development Program of China (2017YFA0402702). We acknowledge the science research grants from the China Manned Space Project with NO. CMS-CSST-2021-B06.
We acknowledge support from the general grants U2031202, 11903083, 11973004 of the National Natural Science Foundation of China. This research has
made use of data provided by the Yaoan High Precision Telescope at Purple Mountain Observatory. This work has made use of data from the European Space Agency (ESA) mission Gaia (\url{https://www.cosmos.esa.int/gaia}), processed by the Gaia Data Processing and Analysis Consortium (DPAC; \url{https://www.cosmos.esa.int/web/gaia/dpac/consortium}). Funding for the DPAC has been provided by national institutions, in particular the institutions participating in the Gaia Multilateral Agreement. 
\end{acknowledgements}

\appendix                  
\section{The conversions between photometric systems}
\label{app:conv}
Since the mass lines and isochrones in ZTF photometric system were not provided by CMD\,3.5 \citep{2017ApJ...835...77M,2019MNRAS.485.5666P}, those in PS1 \citep{2016arXiv161205560C} photometric system provided by CMD\,3.5 are used in Section \ref{sec:sel}. To correct the stellar magnitudes from ZTF to PS1 photometric systems, the conversions between the two photometric system derived from \citet{2020RNAAS...4...38M} are used. The conversions are as follows:
\begin{eqnarray}
  g_{\rm ZTF} - g_{\rm PS1} & = 0.055(g_{\rm PS1} - r_{\rm PS1}) - 0.012, \nonumber\\
  r_{\rm ZTF} - r_{\rm PS1} & = -0.087(g_{\rm PS1} - r_{\rm PS1}) - 0.0035 \nonumber
\end{eqnarray}
and were converted to :
\begin{eqnarray}
  g_{\rm ZTF} - g_{\rm PS1} & = 0.048(g_{\rm ZTF} - r_{\rm ZTF}) - 0.012, \nonumber\\
  r_{\rm ZTF} - r_{\rm PS1} & = -0.076(g_{\rm ZTF} - r_{\rm ZTF}) - 0.0042. \nonumber
\end{eqnarray}

\section{The PHOEBE Model Result}
\label{app:emcee}
The posteriors and uncertainties for the physical parameters of sources 1 and 2 are presented in Figure \ref{fig:ob1} and Figure \ref{fig:ob2}, respectively.
\setcounter{figure}{0}
\renewcommand{\thefigure}{A\arabic{figure}}
\begin{figure}[!ht]
 \centering
          \includegraphics[width=.99\textwidth]{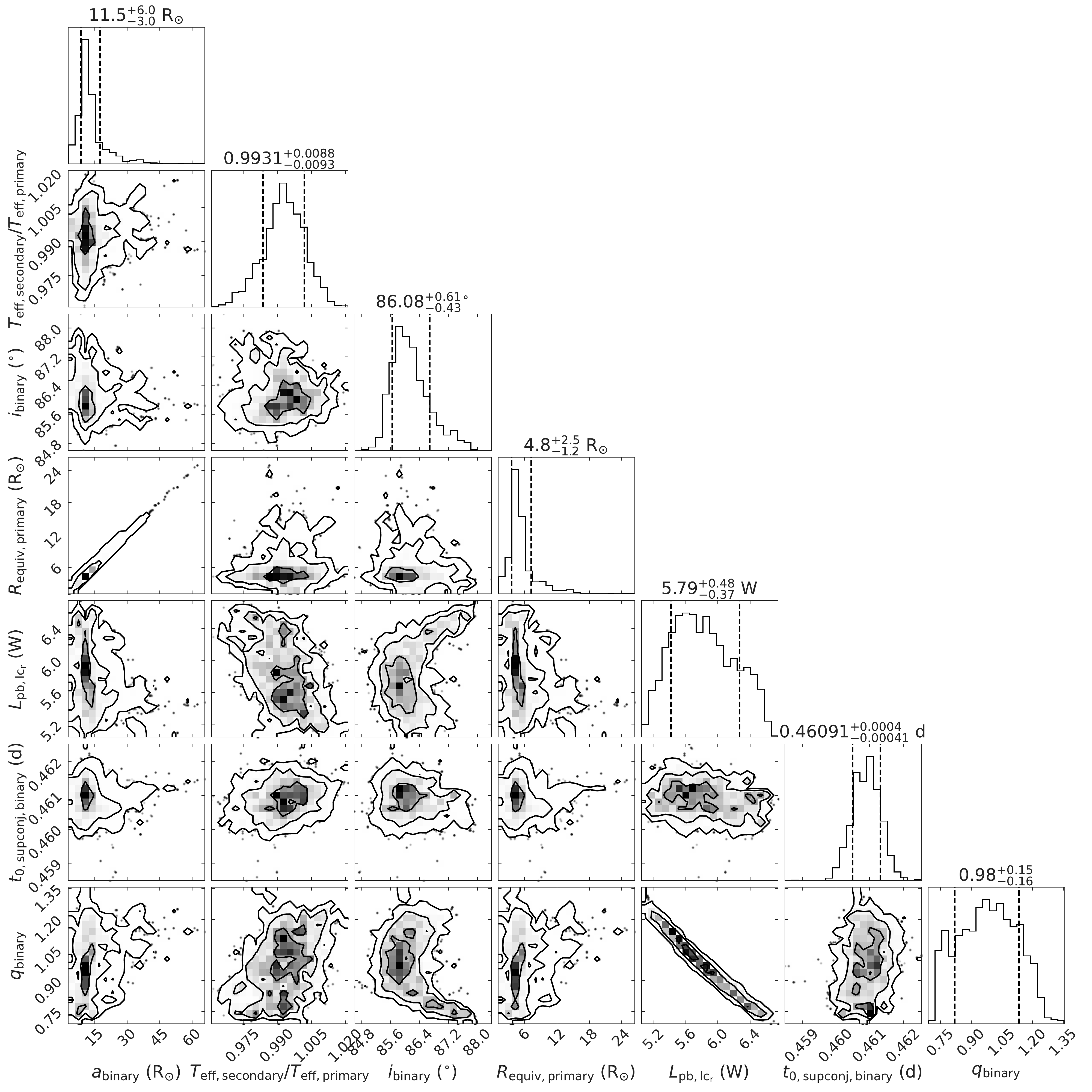}
      \caption{The posterior distributions for the physical parameters of source 1. The dashed lines in histogram plots denote 68\% (1$\sigma$) confidence ranges. The contours in the probability density indicate correspond to 1.0, 2.0, and 3.0$\sigma$ confidence intervals.}\label{fig:ob1}
\end{figure}

\begin{figure}[!ht]
 \centering
          \includegraphics[width=.99\textwidth]{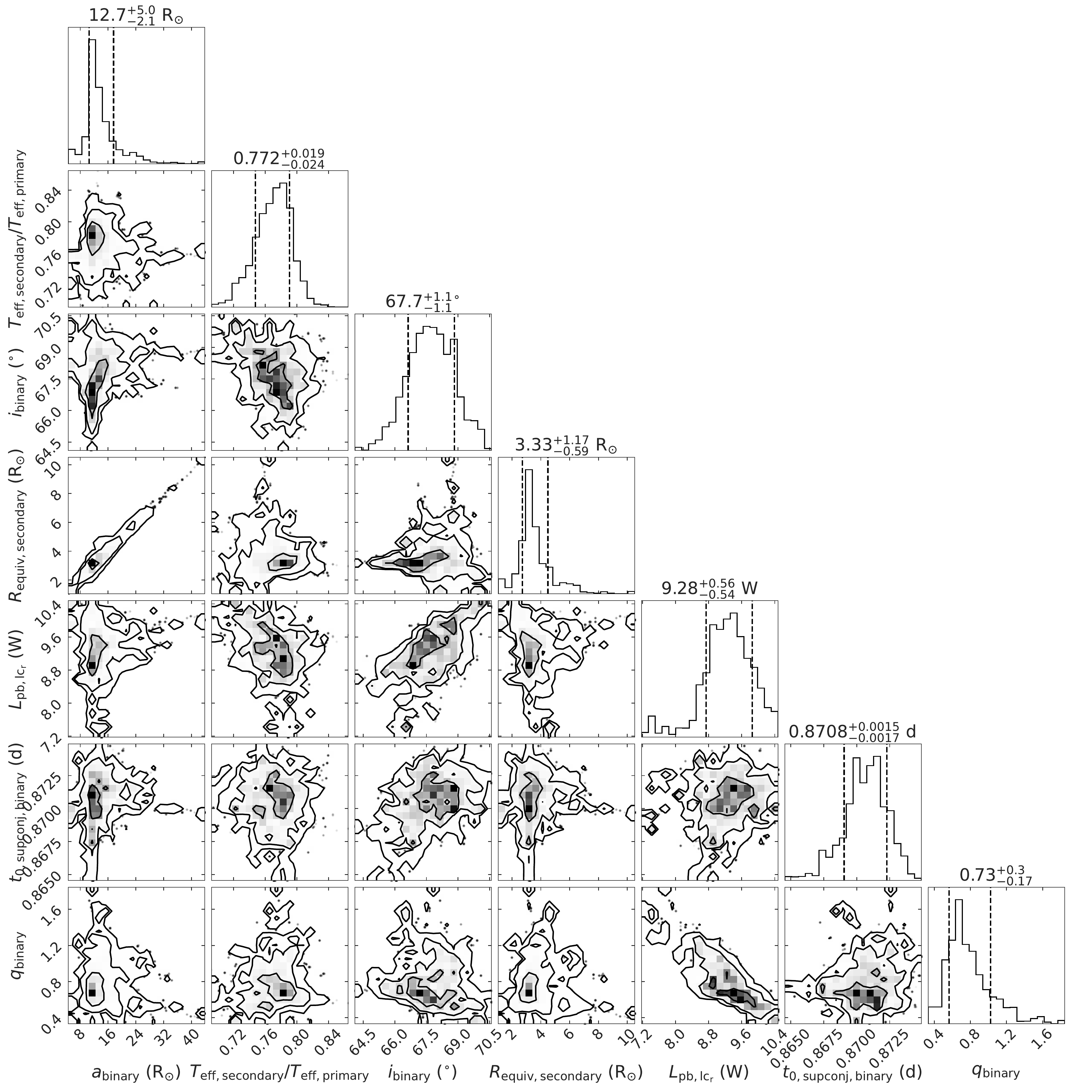}
      \caption{Same as Fig. \ref{fig:ob1}, but for the parameters of source 2.}\label{fig:ob2}
\end{figure}
  
\bibliographystyle{raa}
\bibliography{bibtex}
\end{CJK*}
\end{document}